# Ultraviolet Spectroscopy of Asteroid (4) Vesta

Jian-Yang Li[a,*], Dennis Bodewits[a], Lori M. Feaga[a], Wayne Landsman[b], Michael F. A'Hearn[a], Max J. Mutchler[c], Christopher T. Russell[d], Lucy A. McFadden[e], Carol A. Raymond[f]

[a] Department of Astronomy, University of Maryland, College Park, MD 20742, United States

[b] Adnet Systems, NASA Goddard Space Flight Center, Greenbelt, MD 20771, United States

[c] Space Telescope Science Institute, Baltimore, MD 21218, United States

[d] IGPP & ESS, University of California at Los Angeles, CA 90095, United States

[e] NASA Goddard Space Flight Center, Greenbelt, MD 20771, United States

[f] Jet Propulsion Laboratory, California Institute of Technology, Pasadena, CA 91109, United States

**Proposed Running Head:**

UV spectroscopy of Vesta

**Editorial correspondence to:**

Jian-Yang Li

Department of Astronomy

University of Maryland

College Park, MD 20742-2421

USA

Tel: 301-405-2103

Fax: 301-405-3538

Email: jyli@astro.umd.edu

**Abstract**

We report a comprehensive review of the UV-visible spectrum and rotational lightcurve of Vesta combining new observations by Hubble Space Telescope and Swift Gamma-ray Burst Observatory with archival International Ultraviolet Explorer observations. The geometric albedos of Vesta from 220 nm to 953 nm are derived by carefully comparing these observations from various instruments at different times and observing geometries. Vesta has a rotationally averaged geometric albedo of 0.09 at 250 nm, 0.14 at 300 nm, 0.26 at 373 nm, 0.38 at 673 nm, and 0.30 at 950 nm. The linear spectral slope as measured between 240 and 320 nm in the ultraviolet displays a sharp minimum near a sub-Earth longitude of 20º, and maximum in the eastern hemisphere. This is consistent with the longitudinal distribution of the spectral slope in the visible wavelength. The photometric uncertainty in the ultraviolet is ~20%, and in the visible wavelengths it is better than 10%. The amplitude of Vesta's rotational lightcurves is ~10% throughout the range of wavelengths we observed, but is smaller at 950 nm (~6%) near the 1-µm band center. Contrary to earlier reports, we found no evidence for any difference between the phasing of the ultraviolet and visible/near-infrared lightcurves with respect to sub-Earth longitude. Vesta's average spectrum between 220 and 950 nm can well be described by measured reflectance spectra of fine particle howardite-like materials of basaltic achondrite meteorites. Combining this with the in-phase behavior of the ultraviolet, visible, and near-infrared lightcurves, and the spectral slopes with respect to the rotational phase, we conclude that there is no global ultraviolet/visible reversal on Vesta. Consequently, this implies a lack of global space weathering on Vesta, as previously inferred from visible-near-infrared data.

1. Introduction

Asteroids are generally considered both building blocks of and debris left over from the formation of planetary systems. Knowledge of their composition and mineralogy is mostly collected through spectroscopic observations and comparison with similar reflectance spectroscopic studies of meteorites under laboratory conditions. Most spectral reflectance studies have focused on optical and near-infrared (NIR) wavelengths, yet recent laboratory measurements, space missions, and remote observations with Hubble Space Telescope (HST) demonstrate that asteroidal ultraviolet (UV) spectra contain a wealth of diagnostic information that is yet to be fully understood. The existence of a UV absorption band in the spectrum of Ceres (Li et al., 2009), and the absence of one in the spectrum of Šteins (A'Hearn et al., 2010; Keller et al., 2010) are indicative of substantial spectral differences among different asteroids at these wavelengths. Recent laboratory measurements of the UV spectra of terrestrial minerals, meteorite samples, and frosts of volatiles have shown many potentially diagnostic absorption features in the UV region (see e.g. Cloutis et al. 2008; Hendrix & Vilas 2006). Space weathering is believed to lower the albedo in the visible, subdue the absorption features in the visible and NIR wavelengths (e.g., Chapman, 2004), and flatten out the UV region resulting in brightening and bluing in the UV relative to fresh materials (Hendrix and Vilas, 2006).

The International Ultraviolet Explorer (IUE) telescope collected the first set of UV reflectance spectra of about 50 asteroids in the 1980's (Butterworth et al., 1980; Butterworth and Meadows, 1985). In addition to these IUE observations, only a few scattered observations of some asteroids in the UV exist. The asteroids (2867) Šteins and (21) Lutetia were observed with HST and with the ALICE instrument during the Rosetta spacecraft's flybys (A'Hearn et al., 2010;

Weaver et al., 2010; Stern et al., 2011). Additionally, HST acquired the UV spectrum of (1) Ceres (Parker et al., 2002; Li et al., 2006; 2009), and (596) Scheila was observed with the Swift Gamma-ray Burst Observatory (Bodewits et al., 2011a). In the mean time, laboratory measurements of the UV reflectance spectra of meteorites and candidate asteroid composition minerals have gradually been collected (e.g., Wagner et al., 1987; Cloutis et al., 2008). Despite all the development in laboratory work and UV observations of asteroids, our understanding of asteroid reflectance spectra in this wavelength region is still in its infancy due to limited data in both sensitivity and spectral resolution and equally limited laboratory studies.

As the second largest asteroid by mass, Vesta has drawn the attention of planetary scientists for a long time. It was identified as the most probable source of a class of meteorites, the abundant howardite-eucrite-diogenites (HEDs) collection (McCord et al., 1970). Vesta is recognized as the largest differentiated asteroid in the solar system that has a basaltic surface, indicating it underwent a thermal evolution similar to terrestrial planets in its early history (e.g., Keil, 2002; Pieters et al., 2005). Its unique position in the asteroid population secured its scientific significance to understanding the formation and evolution of terrestrial planets. This is the fundamental motivation of NASA's Dawn mission, which is currently at Vesta for a year long rendezvous. In orbit around it, Dawn will perform a detailed survey of its surface characteristics, internal structure, and elemental abundance targeted to understanding its formation and evolution (Russell et al., 2007).

Rotationally resolved IUE observations showed single-peaked rotational lightcurves in the UV from 240 nm to 320 nm (Festou et al., 1991; Hendrix et al., 2003). The UV albedo of Vesta is less than 0.10, much lower than its visible albedo of ~0.37 (Schevchenko and Tedesco, 2006). While the close analogy between Vesta and HED meteorites at optical and NIR wavelengths has been interpreted as indicating the absence of global-scale space weathering (e.g., Keil, 2002; Pieters et al., 2005), Hendrix et al. (2003) proposed that Vesta's UV spectra and the reversal of lightcurve maximum and minimum in the UV do indicate relatively minor amounts of space weathering.

Vesta has not been studied in the UV since the early IUE observations, now over 30 years ago. With the arrival of Dawn at Vesta in July 2011, better knowledge of this asteroid in the UV becomes necessary to provide context for the interpretation of the spectral data from the Dawn mission. The study of Vesta in the UV provides additional data to constrain knowledge of Vesta's surface mineralogy and its formation history. Of particular interest is what the UV spectral region can reveal about the presence or absence of space weathering on Vesta.

We hereby report observations of Vesta with HST's Wide Field Camera 3 (WFC3) and the UV-Optical Telescope (UVOT) onboard Swift, a multi-spectral gamma-ray burst observatory. We will describe our observations and data reduction and measurements in the next section (Section 2), then discuss the spectrophotometry of Vesta in the UV and at visible wavelengths by combining the HST/WFC3, Swift/UVOT, and IUE observations in Section 3. Section 4 includes a simple mixing model invoked to explain the observed UV-visible spectrum of Vesta, and a discussion about space weathering based on our observations. The conclusions are summarized in Section 5.

## 2. Observations and data reduction

*2.1 Hubble Space Telescope*

### 2.1.1 Photometry

HST observed Vesta in February 2010 with the WFC3 through seven filters centered from 225 nm to 953 nm, as summarized in Table 1. The exposures through filters F373N, F469N, F673N, and F953N cover one full rotation, while the images through F225W, F275W, and F336W only cover one rotational phase at sub-Earth longitude of ~250º. In our calculations of Vesta's longitude/latitude, we adopted the pole orientation of Vesta revised by Li et al. (2011), which represents a change of about 4º from those previously reported by Thomas et al. (1997) and Drummond and Christou (2008). This minor change in the pole orientation causes a shift of only a few degrees in the sub-Earth and sub-solar coordinates on Vesta, and therefore does not affect the comparisons of our work with all previous works based on the pole determined by Thomas et al. (1997).

Starting from the pipeline-calibrated images delivered by HST, we measured the total flux from Vesta in all images within a circular aperture of 40 pixels (1.6") in diameter. The sky background was typically a few counts, negligible compared to more than 1000 counts from the disk of Vesta. The correction for aperture size due to the point-spread-function (PSF) for the extended disk of Vesta (0.54" in diameter for our observation) is different from that for a point source. We extracted the aperture photometry of Vesta with a series of diameters from 0.8" to 8.0", as well as for the model PSF generated by TinyTIM (Krist and Hook, 2004). Assuming that an 8.0" aperture (much larger than the angular size of Vesta) captures more than 99% of the flux, we normalized the energy enclosure curve of Vesta with that of the PSF at 8.0" aperture, and derived the PSF correction for 1.6" aperture photometry of Vesta. The correction factors range from 1.081 through F225W filter to 1.063 through F953N filter (Table 1). The red leak for WFC3 in the F225W filter is less than 0.3% for a solar type star for flux longward of 400 nm (Rajan et al., 2010). We estimated the red leak of F225W for Vesta by modulating a solar spectrum E490 (E-490-00, American Society for Testing and Materials Airmass 0 reference spectrum, DOI: 10.1520/E0490-00AR06) with a preliminary reflectance spectrum of Vesta. We constructed the latter by combining our photometric measurements in the visible wavelengths with archival spectra from the SMASS II and IUE and folded it through the filter transmission curves. The fraction of the flux longward of 400 nm is estimated to be ~2.8% for Vesta, comparable to that of a star with an effective temperature of 4000 K. We corrected for the red leak in the F225W photometry by subtracting 2.8% flux from the total measured flux. The red leak for F275W and F336W filters at this effective temperature is negligible, amounting only to 0.3% and 0.01%, respectively.

*2.1.2 Albedo*

The geometric albedo of Vesta, $p$, is calculated as,

$$p = \frac{F_V \, r_\odot^2 \, \Delta^2}{F_\odot \, R^2 f(\alpha)}$$

where $F_V$ is the measured flux of Vesta; $F_\odot$ is solar flux at 1 AU as measured through the same filter; $r_\odot$ and $\Delta$ are the heliocentric distance and geocentric distance of Vesta at the time of observation; $R$ is the equivalent diameter (the diameter of a spherical body with the same cross-sectional area) of Vesta, and $f(\alpha)$ is the phase function of Vesta. We took the high-resolution solar spectrum E490 and the synthetic filter transmission curves provided by HST (Dressel, 2010) to calculate the equivalent solar flux density through each filter we used:

$$F_\odot = \frac{\int S(\lambda)\, T(\lambda)\, \lambda\, d\lambda}{\int T(\lambda)\, \lambda\, d\lambda}$$

where $S(\lambda)$ is solar spectrum; and $T(\lambda)$ is the composite filter transmission curve including CCD quantum efficiency and telescope optics. The calculated values of $F_\odot$ for all seven filters are listed in Table 1.

A phase function correction is critical to deriving the geometric albedo, but different phase functions may be used. For example, Festou et al. (1991) used a phase function of a constant slope of 0.036/deg for the IUE observation at 17º phase angle, while Roettger and Buratti (1994) used a Hapke phase function for average S-type asteroids (Helfenstein and Veverka, 1989) for the same data. Based on ground-based photometry data of Vesta in the *V*-band (Lagerkvist et al., 1995), we applied an IAU HG phase function model (Bowell et al., 1989) with a G parameter of 0.27 (Fornasier et al., 2011), and assumed a constant phase function across the whole range of wavelengths of our observations. Phase reddening is expected to result in a steeper phase slope at shorter wavelengths (Reddy et al., 2011). However, there is currently no available model to describe phase reddening, either physical or empirical. The sparse phase angle coverage and uncertainties introduced by using measurements obtained with different instruments prevent us from deriving the phase reddening function here. We could therefore not correct our data for this effect. We will return to this issue in Section 2.4 where we compare observations from different instruments. The cross-sectional area of Vesta at the sub-Earth latitude of the HST observation (~20º) was calculated using the shape model as well as the pole orientation (Li et al., 2011) to have an equivalent diameter of 518 km. With only ~3% difference between the long-axis and the intermediate axis, we ignored the change in the long-axis in the projected disk of Vesta, and used a constant cross-sectional area in our calculation.

In order to best compare the broadband albedo with spectroscopic measurements, we invoke the concept of "equivalent wavelength". Since the solar flux drops by two orders of magnitude between 400 nm and 200 nm, the broadband geometric albedo measured in this range depends significantly on the transmission profile of the filter and the illumination source used, and could be significantly different from the central wavelengths of filters. The measured albedo is weighted more towards longer wavelengths where solar flux is stronger. Equivalent wavelength of broadband albedo is defined as the transmitted source flux weighted average wavelength of the filter transmission:

$$\lambda_E = \frac{\int S(\lambda)\, T(\lambda)\, \lambda\, d\lambda}{\int T(\lambda)\, d\lambda}$$

where $S(\lambda)$ is the solar spectrum, and $T(\lambda)$ is the filter transmission curve. Since the red leak of WFC3 broadband filters in the UV is only up to 3% for F225W and negligible for longer wavelengths, its effect on the equivalent wavelength is ignored. The equivalent wavelengths of all the filters we used are listed in Table 1. For narrow band filters, the equivalent wavelengths are almost independent of the source spectrum.

*2.2 Swift*

*2.2.1   Broadband photometry*

Swift observed Vesta on April 5 and 20, 2011 UT (Table 1). Our observations used the UV-Optical Telescope (UVOT) that provides a 17 × 17 arcminute field of view with a plate scale of 1 arcsec/pixel. The PSF of UVOT has a FWHM of 2.5" (Poole et al., 2008). We used the UVW2 and UVM2 filters, and the UV grism to acquire low-resolution ($\lambda/\delta\lambda$ = 100) spectroscopy between 200 and 330 nm. The grism was operated in 'clocked mode' to suppress background stars and the dispersion axis was oriented at a position angle of ~260°. The asteroid was not tracked, and the proper motion is ~10" over our ~700 s grism exposures, ~2" for UVM2 exposures, and ~0.3" for UVW2 exposures, with varying effects on the photometry as discussed below.

We measured the broadband photometry through UVM2 and UVW2 from the standard 5" radius aperture (Poole et al., 2008). Compared to the size of the PSF, the smearing in the images is small and barely visible in UVM2 images, and the effect on photometry is negligible. The brightness of Vesta results in significant coincidence loss (Poole et al., 2008), which we estimated to be 0.7 mag at UVW2 and 0.06 mag at UVM2 and corrected for in our photometric measurements. While UVM2 only suffers a minimal amount of red leak for solar spectrum, UVW2 has a significant red leak. We took the filter transmission profile (Breeveld et al., 2011), and multiplied it with both solar spectrum (E490) and the solar spectrum reddened by a trial Vesta spectrum in the UV-visible wavelength (based on our HST measurements and IUE observations) to estimate the fraction of flux coming from long wavelength. For UVM2, about 0.6% flux comes from wavelength longer than 400 nm for a solar spectrum, and 2.6% for a Vesta-reddened solar spectrum; for UVW2, it is about 25% for a solar spectrum, and 50% for a Vesta-reddened spectrum. Based on those factors, we removed the out-of-band flux from the total Vesta's flux measured from imaging photometry, and then only used the in-band solar flux to derive the geometric albedo for the two filters we used. A red leak also affects the effective wavelength. Following the same procedure described in the previous section, and using the solar spectrum E490, we calculated the expected solar flux through the UVOT filters, as well as the effective wavelengths listed in Table 1. With the effort to remove the red leak from the filters using the fraction of flux below 400 nm, it is reasonable to set the flux to zero beyond 400 nm in calculating the effective wavelengths.

The photometric results are subject to several possible systematic uncertainties. The absolute calibration of both UVW2 and UVM2 is accurate to within 0.03 magnitudes. Shot noise and background subtraction uncertainty, as well as the aperture and coincidence loss correction errors, are all included in the listed uncertainties. For the red leak, we estimated the uncertainty at 10% after correction. All sources of errors are combined assuming they occur in quadrature.

*2.2.2 Grism spectroscopy*

We extracted the asteroid's spectrum from a rectangular region 13 pixels wide, centered on its position. Swift did not track the asteroid, and the combined smearing due to the orbital parallax and apparent motion of Vesta is ~20 pixel along the dispersion axis, and ~7 pixels in the perpendicular direction. The PSF along the spatial direction is about ~8 pixels in FWHM, and we found that the smearing affects the photometry at a 4% level. The smearing along the dispersion axis effectively degrades the spectral resolution to ~5 nm. We accounted for this in the removal of the solar spectrum by binning Vesta's spectrum in 10 nm intervals, and by smoothing the solar spectrum with a 10 nm boxcar average. The brightness of the asteroid resulted in saturation or significant coincident loss of the spectrum beyond 290 nm. Background stars in the dispersion tracks contaminate two of the four spectra we acquired. We were able to subtract the stellar spectrum from Vesta's spectra obtained by comparison with a grism image where the background star did not fall onto Vesta's dispersion track. The subtraction resulted in a spectrum of Vesta that is in good agreement with the uncontaminated spectra (Fig. 1). The uncertainty of the absolute radiometric scale of Vesta's spectrum is 25% (Bodewits et al., 2011b). The relative pixel-to-pixel noise in the spectrum is typically less than 2% after binning.

The reflectance spectrum and geometric albedos were derived using the methods and solar spectrum discussed in Section 2.1.2. The sub-Earth latitude of Vesta during the Swift observations is similar to that during HST's observations except that it is on the opposite hemisphere. Therefore the same equivalent diameter of Vesta as used for the HST data is used here. The geometric albedo spectra of Vesta measured by Swift are shown in Fig. 1.

*2.3 International Ultraviolet Explorer*

IUE observed Vesta in 1978-1979 (e.g., Butterworth et al., 1980; Roettger and Buratti, 1994). In order to compare the IUE data to our new observations, we re-processed the IUE spectra using the same procedure, the same solar spectrum, and the same phase function in the derivation of geometric albedos as were used for the HST and Swift data.

The flux spectra of Vesta were extracted from all relevant IUE data (NEWSIPS) that are available through the Multimission Archive at STScI (MAST) website. We applied the corrections to all Vesta's spectra to derive absolute flux following Massa and Fitzpatrick (2000) to all Vesta spectra. This correction is in general at a 5% level. We used the same data set that was used by Hendrix et al. (2003). Earlier IUE data are either of poor quality or incomplete, and are not included in our study. The sub-Earth latitude of Vesta during the IUE observations was about 10º North. The difference in the cross-sectional area of Vesta from HST and Swift observation is at a 1% level, and is therefore ignored.

*2.4 Comparisons between three instruments*

Fig. 2 shows the geometric albedo measured by HST, Swift, and IUE at various wavelengths, combined with SMASS II data (Bus and Binzel, 2003). Because SMASS II only contains normalized reflectance spectra, they were scaled to match the average of our HST photometry at 673 nm.

A critical parameter to derive geometric albedos from flux measurements is the assumed equivalent diameter. Assuming an equivalent diameter of 468 km, IRAS observations yielded a geometric albedo of 0.42±0.05 at visible wavelengths (Tedesco et al., 2002). Corrected to the equivalent diameter of 520 km derived from HST observations (Thomas et al., 1997), the IRAS measurement is equivalent to a geometric albedo of 0.34, consistent with our HST measurement (Fig. 2a). Based on an occultation-determined size, Shevchenko and Tedesco (2006) reported a visible geometric albedo of 0.37 for Vesta, consistent with both the adjusted IRAS measurements and our measurements presented here.

Our UV measurements using WFC3 yield geometric albedos of Vesta of 0.12 at 253 nm, 0.15 at 282 nm, and 0.21 at 338 nm. This is almost twice the albedos of 0.065 at 245 nm and 0.075 at 270 nm as previously reported by Festou et al. (1991) and Roettger and Buratti (1994) based on IUE data. These differences can be partially explained by the different phase function corrections and possibly the equivalent size of Vesta we used. Applying the phase function used for the Swift and HST data to IUE data, we found geometric albedos of 0.074 at 245 nm and 0.085 at 270 nm. We assumed here that Festou et al. (1991) and Roettger and Buratti (1994) used an equivalent diameter similar to ours, as this parameter was not reported in their papers. Hendrix et al. (2003) reported the reflectance of Vesta at a phase angle of 17º to be between 0.04 and 0.07 between 240 nm and 320 nm using an equivalent diameter similar to ours. Applying our phase function, Hendrix et al.'s reflectance translates to a range of 0.08 and 0.14 geometric albedo, in much better agreement with our measurements, yet still lower by about 25% (Fig. 2b).

Another factor that complicates the comparisons of observations made under different geometries is phase reddening (Section 2.1.2). While Vesta shows only a slightly redder reflectance spectrum in the visible wavelengths when observed at larger phase angles than at smaller phase angles, laboratory measurements of the reflectance of HED meteorites do show some obvious reddening with phase angle (Reddy et al., 2011). If phase reddening extends into the UV, then it would result in a steeper phase function at shorter wavelengths. The *V*-band phase function correction we used in the UV may result in under-correction for data at higher phase angle than at lower phase angle. This is consistent with the fact that the measurements of UV albedo from Swift (~26º phase) are the lowest, the values from HST data (~5º phase) are the highest, and those from IUE data (~17º phase) are in between. A quantitative estimate based on the formulae proposed by Reddy et al. (2011) shows that at least 1/3 of the difference between Swift and HST observations, and at least 1/10 difference between IUE and HST observations could be accounted for by phase reddening, assuming those formulae are valid in the whole range of wavelengths and range of phase angles of our study including inside the opposition surge (phase angle < a few degrees).

To compare the absolute radiometric calibrations between different instruments, we took one IUE flux spectrum (LWP18950) that is close to the median of the rotational lightcurve, and passed it through the WFC3 F225W and F275W filters to calculate the corresponding broadband fluxes. The predicted WFC3 flux from the IUE spectrum should be good within 25%. After correcting for the observing geometry and phase angle (no phase reddening considered), the total WFC3 broadband fluxes calculated from the IUE spectrum are lower than WFC3 measurement by 20%. This agrees with the previous comparisons between WFC3 and IUE observations (Fig. 2b).

Could the differences in Vesta's UV flux measured using different instruments be due to the variations in the incident solar flux at the time of each observation? The IUE observation was made near a solar maximum in 1990, while the Swift and HST observations were both obtained near a solar minimum. While the solar flux does vary more in the UV than at longer wavelengths within solar cycles, the variation is about 10% at 200 nm, 5% at 250 nm, and less than 1% at wavelengths longer than 300 nm (Lean, 1989). There are no solar spectral data available for the time of IUE observations, but the variations in solar flux from solar maximum to solar minimum suggest that the albedo spectra of Vesta derived from IUE data could overestimate the albedo of Vesta by ~10% at 200 nm, and by ~1% at 300 nm. This can explain why Vesta's UV albedo measured from Swift is 10%-20% lower than those from IUE, and also explains the slight differences in the spectral slopes measured at different epochs (phase reddening being another possible factor, but hard to quantify).

Combining all the possible sources discussed above that may cause this discrepancy, we conclude that the geometric albedo measurements in the UV from three instruments at three epochs and observing geometries are consistent with each other. The observed differences are caused by the combination of phase reddening, solar spectrum, and instrumental differences. Considering the uncertainties of the absolute flux calibration of the Swift grism of ~25% (Bodewits et al., 2011b), of IUE, 15-20% (Nichols and Linsky, 1996; Massa and Fitzpatrick, 2000), and that HST UV measurements were made through broadband filters, we conclude that the derived UV albedos are in good agreement. Given the fact that HST observations were performed at the lowest phase angle of ~5º, and that the absolute flux calibration of broadband photometry should be better than that for spectroscopic observations, we consider that the UV geometric albedos derived from HST observations are the most reliable. In addition, our geometric albedos are consistent with the most recent measurements of Vesta by the OSIRIS cameras onboard the Rosetta spacecraft (Fornasier et al., 2011).

## 3. Results

*3.1 UV-visible reflectance of Vesta*

The UV albedo of Vesta of about 0.08 at 250 nm is comparable to other asteroids. The albedos of asteroids at UV wavelengths are generally low. IUE measured the UV spectra of 45 asteroids of almost all major taxonomies including C, S, V, M, etc. (e.g., Butterworth and Meadows, 1985; Roettger and Buratti, 1994). Only (44) Nysa (E-type) has a high geometric albedo of ~0.28 at 250 nm; all others have geometric albedos below 0.12, and mostly between 0.03-0.08. The most recent measurement of Ceres (G-type) yielded a geometric albedo of ~0.03 near 280 nm (Li et al., 2006; Rivkin et al., 2011), which is consistent with values previously reported from IUE (Butterworth et al., 1980; Butterworth and Meadows, 1985; Roettger and Buratti, 1994). Asteroid Šteins (E-type) has a relatively high geometric albedo of ~0.15 at 250 nm (Keller et al., 2010; A'Hearn et al., 2010). Lutetia, which is specified as an M-type in Tholen's taxonomy but an X-type by Bus and Binzel (2002) for its unusual spectrum and albedo in the visible, has a geometric albedo of ~0.1 at 250 nm (Weaver et al., 2010; Stern et al., 2011). Asteroid (596) Scheila, which showed an ejecta plume in December 2010 that is mostly consistent with an impact, is a T-type asteroid (Bus and Binzel, 2002) and has an albedo of 0.02-0.03 between 200 nm and 300 nm (Bodewits et al., 2011a). The limited UV studies available suggest that the major taxonomic classes likely persist in the near-UV (Roettger and Buratti, 1994).

Although almost all of the asteroids with UV observations have had their geometric albedos and spectra observed in the visible as well, only those mentioned above have either been studied in parallel with both UV and visible spectroscopy or have had their UV and visible spectra systematically discussed in a compilation. Within the whole spectral range from visible into UV wavelengths, asteroids display entirely different spectral absorption characteristics. Vesta has a strong absorption band near 1-μm starting near 750 nm diagnostic of mafic minerals, a linear red slope of about 7-10%/100 nm in the visible outside of the 1-μm band, and then a very steep and almost linear spectrum into UV until 220 nm (Fig. 2). By comparison, Ceres has a flat spectrum through the visible and a steep drop starting from just longer than 400 nm wavelength, leading to a deep (~60%) absorption band centered at 280 nm (Li et al., 2006; 2009). Šteins has a weak red slope of about 7%/100 nm in the visible, and then a gradual drop in albedo from 0.3 at 500 nm to 0.05 at 200 nm, by almost 85% (A'Hearn et al., 2010). Lutetia has a ~2%/100 nm red slope at visible wavelengths, a quick drop from 0.15 at 350 nm to 0.1 at 300 nm, and then stays almost flat down to 150 nm (Weaver et al., 2010; Stern et al., 2011). Asteroid Scheila is very red in the visible with almost a 40%/100 nm slope, and has a possible absorption band of about 15% centered at 350 nm (Bodewits et al., 2011a). The various UV slopes and absorption features in the UV are associated with particular minerals and may contain diagnostic information about the surface compositions of asteroids (e.g., Cloutis et al., 2008; Wagner et al., 1987). It is thus valuable to combine the visible spectroscopy with the UV wavelength data to systematically study the spectral characteristics of a large sample of asteroids.

*3.2 Rotational lightcurve*

Fig. 3a shows the geometric albedo measured from HST/WFC3 and Swift images plotted with respect to sub-Earth longitude; Fig. 3b shows IUE spectra grouped into three wavelength bins and plotted with respect to longitude. The shapes of the four full lightcurves at 373, 469, 673, and 953 nm are consistent with those reported before (e.g., Blanco and Catalano, 1979). The UV lightcurves have similar shapes as those in visible wavelengths. All lightcurves are dominated by albedo rather than cross-section variations, with a darker western hemisphere and a brighter eastern hemisphere (Binzel et al., 1997; Li et al., 2010). The amplitudes of the lightcurves are ~10% in 373 nm to 673 nm, ~6% at 953 nm near the center of the 1-μm absorption band characteristic of its basaltic surface, and ~10% in the UV. The lightcurve maxima in the four full lightcurves from WFC3 occur at almost the same longitude near 100º. The lightcurve minimum appears to be flat between 300º and 0º in 953 nm, and occurs near ~300º or slightly to the west for 673 nm and shorter wavelengths into the UV. The HST snapshot geometric albedo measurements at 336 nm, 275 nm, and 225 nm correspond to sub-Earth longitudes of 280º to 290º, near the lightcurve minima in other wavelengths.

Based on the IUE observations, Hendrix et al. (2003) reported the exchange of Vesta's lightcurve maximum and minimum from visible wavelength to UV, and attributed this effect to space weathering. However, the sub-Earth (west) longitudes we calculated from the reported IUE observing times are consistently greater than those listed in Table 1 in Hendrix et al. (2003) by ~133º. This shifts the lightcurve minimum to about 290º E, and the maximum to about 120º E, *consistent with the lightcurves in the visible wavelengths*. Festou et al. (1991) phased the same IUE UV spectra and simultaneous photometric data in near-*B* band obtained with the FES camera onboard IUE, and reported that the UV lightcurve and near-*B* band lightcurve are slightly out-of-phase by up to 20º. Taking the coordinate system defined by Li et al. (2011), the sub-Earth longitude at the zero rotational phase in Festou et al. is 85º, putting both the near-*B* band lightcurve and the UV lightcurve in good agreement with the lightcurves we observed from WFC3. The amplitude of the Festou et al. near-*B* lightcurve of ~11% is the same as the amplitude of our lightcurve at 469 nm. In addition, our Swift photometric data (Fig. 3a), although sparse, are not consistent with a reversal of maximum and minimum between UV and visible lightcurves. The rotational period of Vesta was measured to a precision better than 0.004 seconds (Drummond et al., 1998), resulting in an uncertainty of only ~2º in longitude over 20 years. We therefore conclude that, although Vesta's lightcurves at wavelengths between 240 nm and 953 nm appear to be slightly shifted in phase, there is no evidence for a complete lightcurve reversal at those wavelengths. The geometric albedos of Vesta at 225 nm, 275 nm, and 336 nm derived from our HST observations were obtained near the minimum of the rotational lightcurves of Vesta, and the maximum brightnesses at those wavelengths are about 10% higher.

*3.3 Rotational variation of the UV-spectrum*

To further explore rotational variations of the UV spectrum of Vesta, we measured the UV spectral slope of Vesta from IUE data within 240-320 nm by a linear fit (Fig. 4). The UV slope appears to have a dip (relatively blue) near 10º in longitude, and is relatively high (red) in other longitudes, with the highest values concentrated in the eastern hemisphere from 80º to 160º in longitude. Color ratio maps of 673/439 nm shows that the reddest area is located between 60º and 180º in longitude; that the western hemisphere from 180º to 280º is slightly redder than the average surface; and that the bluest area is concentrated between 0º and 40º (Li et al., 2010). This is entirely consistent with the UV slope (240 – 320 nm) we derived from IUE data, indicating its variation with respect to longitude most likely real. This suggests that the same spectral slope variations of Vesta with respect to longitude in the visible continues into UV wavelengths, and the spectral slope map of 673/439 nm (Li et al., 2010) in visible wavelengths is probably still valid in the UV.

While the visible color slope map of Vesta (Li et al., 2010) shows about 15% variation at a 39 km/pixel scale, Fig. 4 shows that at a hemispheric scale, the UV slope changes at least 30% from minimum to maximum. This is evidence that the color variation on Vesta in the UV is larger than that in the visible. IUE lightcurves also show comparable or slightly higher amplitudes than those in longer wavelengths (Fig. 3b).

## 4. Discussion

*4.1 Spectral mixing model*

Vesta's prominent spectral features in the wavelength regime studied here include a sharp absorption edge below 400 nm which is characteristic of charge transfer absorptions and a broad absorption band near 950 nm attributed to electronic absorptions in ferrous iron in pyroxene (McCord et al. 1970; Wagner et al. 1987). From the rotational variation of Vesta's 1-μm and 2-μm mafic bands, including their band centers and the band area ratio, it is generally concluded that the surface of Vesta is mostly covered by brecciated howardite with eucritic and diogenitic minerals concentrated in the western hemisphere and eastern hemisphere, respectively (Gaffey, 1997; Binzel et al., 1997). The spectral characteristics of the UV region are shaped by the combination of multiple charge transfer absorptions; specifically, absorptions due to Fe-O, $Fe^{2+}$, $Ti^{4+}$, Fe-Ti, and $Fe^{2+}$-O, etc. comprise the UV absorption edge (Wagner et al., 1987; Cloutis et al., 2008). While our data do not have the spectral resolution near the 1-μm band to permit us to fit a unique pyroxene abundance model, they cover both the UV and visible wavelengths. We examined the laboratory UV-visible spectra of various minerals and the existing mixing model for Vesta over the whole spectral range against our data.

Variations in Vesta's lightcurve are dominated by compositional variation rather than shape. We adopted a spectrum of Vesta using the midpoints of the UV and visible photometric variations. The lightcurve maxima and minima do not have an obvious shift across the wavelength regime of our study (Fig. 3), and a model with lightcurve midpoints should represent the average composition of Vesta. We approached our fitting analysis by choosing the most likely compositional candidates available in the laboratory reflectance spectral databases.

Spectra were retrieved from Wagner et al. (1987) for UV wavelengths and from RELAB for visible wavelengths for howardites (Bununu and EET87503, respectively) and diogenite mineral powders (Johnstown for both). Wagner et al. acquired UV spectra of several minerals for a single particle size with the powder sieved to <150 μm. In contrast, RELAB acquired spectra for several particle sizes. In order to investigate the spectral effects of grain size from 300 nm through 1000 nm, we scaled the RELAB reflectances to Vesta's geometric albedo derived from the mid-point of the HST observations at 469 nm. In Fig. 5a, the small-grain howardite spectrum is a uniform, linear mixture of fine powder, both wet and dry sieved to 25 microns. The large-grain spectrum is for powder sieved to 250 microns. The diogenite powders are sieved to 25 and 75 microns, respectively. The finer grain material fits better in both wavelength regimes, consistent with previous studies (Hiroi et al., 1994). This is consistent with previous conclusions that the surface of Vesta is dominated by brecciated howardite and polymict eucrite (Gaffey, 1997). Both howardite with grain size > 25 μm and diogenite resulted in deeper 1-μm bands than observed, although their UV slopes match the observations well. Weathered diogenite samples, which have a weaker 1-μm band than the unweathered diogenite, could fit the 1-μm band but not the spectral slope in the UV.

Because the finer grain spectra from RELAB do not extend down to 200 nm, we created a composite reflectance spectrum of howardite using the fine grain RELAB reflectances above 320 nm, again scaled to the HST geometric albedo mid-point at 469 nm, and the UV reflectance spectra published by Wagner et al. (1987) below 320 nm, scaled to the IUE geometric albedo at 320 nm, so that the fit at long and short wavelengths was optimized. The same thing was done for diogenite. By linearly mixing 20% of the composite diogenite with 80% of the composite howardite, only slight differences are found as compared to 100% fine particle howardite (Fig.

5b). The differences are well within the error bars and absolute calibration of the instruments in the UV. The strength of the 1-µm band puts a strong constraint on the relative abundance of diogenite, with more diogenite resulting in too strong an absorption to match Vesta's visible spectrum.

We re-examined the Hendrix et al.'s (2003) fit to the IUE data with a linear combination of Wagner et al.'s (1987) sample powders (contaminated by large grains). The best fit for the western hemisphere (the original eastern hemisphere in Hendrix et al. (2003)) was a linear combination of 54% howardite, 36% diogenite, and 10% fayalite. For the eastern hemisphere, their best fit included 90% howardite and 10% diogenite. In Fig. 5b, we plot these same linear combinations of reflectances, scaled to the HST data point at 469 nm, with the HST, Swift and IUE geometric albedos. In the UV region, their mixing model fits our data in the UV very well. The discrepancy between their model and our data in the 1-µm region is due likely to the large grain size of the measured meteorites' spectra rather than their modeled compositions. We note that we limited this study to UV and visible wavelengths only. Complete compositional information has to be extracted from all wavelength regions including both the 1-µm and the 2-µm bands. Our spectral model is meant to be a consistency check rather than a unique solution of the composition on Vesta.

In conclusion, a single, fine-grain howardite (grain size <25 µm) fits our Vesta spectrum from 220 nm to 950 nm, including the spectral slope in the UV and in the visible and the short wavelength side of the 1-µm band. Space weathering on a global scale is not required here to match the whole UV-visible spectrum. Although the spectral uncertainty in the UV and the spectral resolution in the visible do not allow us to fit the exact fraction of possible pyroxene components or to study the rotational variations of surface composition, it is evident that the UV region provides additional constraints on the interpretations of the reflectance spectrum of Vesta. The different characteristics of UV spectra of the asteroids that have been intensively studied recently, including Ceres, Šteins, Lutetia, and Scheila, suggest the necessity to systematically study the UV absorption features of asteroids, as well as the laboratory measurements of meteorite samples and terrestrial minerals and ices.

*4.2 Space weathering*

Space weathering is the alteration of the surfaces of airless bodies when exposed to space environment, changing their optical properties by darkening of albedo, reddening of spectral slope, and weakening of absorption features in the visible and NIR. It was originally invoked to explain the spectral differences between lunar soils and rocks, and between returned lunar samples and telescopic observations of the sampling sites on the Moon (e.g., Chapman et al., 2004, and references therein). The more weathered lunar soil samples display much redder spectra than freshly crushed lunar rock samples. Vapor deposition of nanophase iron on the surfaces of regolith grains induced by micrometeorite impact, solar wind particle bombardment and implantation are generally considered to be the primary causes of the change in the optical properties of weathered surfaces (e.g., Pieters et al., 2000). The flybys of two asteroids by the Galileo spacecraft, (951) Gaspra and (243) Ida, the rendezvous of the NEAR Shoemaker spacecraft with a near Earth asteroid, (433) Eros, and the sample return mission Hayabusa to (25143) Itokawa, all being S-type asteroids, provided direct evidence of weathering on this class of asteroids (Chapman et al., 1996; Clark et al., 2001; Naguchi et al., 2011). Based on the analysis of IUE and HST observations of S-type asteroids and laboratory measurements of lunar samples in the UV, Hendrix and Vilas (2006) showed that space weathering manifests itself in the UV as bluing the spectrum as opposed to reddening in the visible-NIR. They also showed that the UV spectrum is more sensitive to space weathering than the visible-NIR spectrum.

Vesta's basaltic surface is thought to be similar to that of the Moon, which is severely weathered. Therefore the lack of any evidence of space weathering on Vesta as suggested by the nearly perfect match between combinations of HED meteorite spectra and Vesta's spectrum (McCord et al., 1970) is always surprising. Both observations of S-type asteroids in the inner main belt and laboratory ion irradiation experiments with meteorite samples suggested that the surface of Vesta should be significantly darkened and reddened with respect to HEDs in $10^6$ yr, a relatively short time scale compared to impact resurfacing time scale (e.g., Nesvorny et al., 2005; Vernazza et al., 2006). Several theories have been proposed to explain the apparent lack of space weathering on Vesta, including resurfacing by a recent large impact, magnetic field shielding, compositional and environmental differences between Vesta's surface and the lunar surface, etc., all with their own deficiencies or aspects that are hard to prove (e.g., see Chapman et al., 2004). Based on the albedo and color ratio maps derived from HST observations, Li et al. (2010) proposed that a few small isolated areas on Vesta might show signs of weathered/fresh surfaces.

As discussed in the previous sections, Vesta's lightcurve amplitude is almost constant in the UV-visible region outside of the 1-µm band; the phasing of its lightcurve maximum and minimum with respect to longitude remains nearly unchanged throughout the UV-visible region; and the longitudinal variation of its UV spectral slope is entirely consistent with that in the visible. What do these results imply about the space weathering on Vesta?

If space weathering on Vesta behaves like on S-type asteroids, too small to be detected in the visible but evident in UV, then its UV slope would show a slight bluing (less red slope) relative to its visible slope (Hendrix and Vilas, 2006). In that case, a smaller color variation (lightcurve amplitude) is expected in the UV than in the visible. Li et al. (2010) argued that the color heterogeneity on Vesta might be mostly related to composition rather than maturity because of the correlation between a relatively redder spectral slope and the relatively stronger 1-µm band depths in most of the surface, except for a few small areas that are either the bluest or the reddest. They also suggested that the bluest area near 20º longitude and -40º latitude might consist of fresh, newly exposed materials; and the extremely red area near 100º longitude and -20º latitude might be the most weathered terrain. Since space weathering makes UV slopes bluer (Hendrix and Vilas, 2006), we would see a relatively steep spectral slope at 20º longitude in the UV. But this is the opposite of what we observe in Fig. 4. Therefore, we conclude that our observations are not consistent with space weathering in the UV-visible region. It is of note that the UV space weathering study was based on the comparisons between S-type asteroids and ordinary chondrites (Hendrix and Vilas, 2006). Any possibly minor or local space weathering on Vesta could manifest itself in a completely different way from that on S-type asteroids, or the Moon, due possibly to different compositions and/or geological histories.

## 5. Conclusions

We have presented a comprehensive review of new and existing observations of Vesta in the regime $0.2 - 1\mu m$. The findings can be summarized as follows:

The geometric albedo derived from photometric and spectroscopic measurements by Swift/UVOT, HST, and IUE between 220 nm and 950 nm in wavelengths are consistent with each other, and consistent with previous measurements in the visible and in the UV, as well as the most recent measurements from the OSIRIS cameras onboard the Rosetta spacecraft (Fornasier et al., 2011).

The rotationally averaged geometric albedo of Vesta is 0.09 at 250 nm, 0.14 at 300 nm, 0.26 at 373 nm, 0.38 at 673 nm, and 0.30 at 950 nm. This is in line with UV albedos measured for other asteroids in the UV region, typically below 0.1. The UV spectrum of Vesta appears to have different characteristics from other asteroids that were observed intensively in the UV recently, including Ceres, Šteins, Lutetia, and Scheila.

Reanalysis of IUE spectra, combined with our rotational lightcurves in the visible, shows that the UV lightcurve is consistent with the visible lightcurve. The amplitude of Vesta's rotational lightcurve remains ~10% throughout UV-visible wavelengths, and decreases to ~6% near the 1-μm band center. The lightcurve maximum occurs in the eastern hemisphere and the minimum occurs in the western hemisphere. The linear spectral slope of Vesta between 240 nm and 320 nm displays a correlation with longitude, with the bluest area concentrated near 20º longitude, the reddest area in the eastern hemisphere, and moderately red area in the western hemisphere. This is entirely consistent with the behavior of the spectral slope in the visible wavelengths from 439 nm to 673 nm, suggesting that the spectral slope map reported by Li et al. (2010) might remain similar into the UV.

We found a nearly constant amplitude of Vesta's rotational lightcurve in the UV through visible wavelengths, with almost simultaneous lightcurve maximum and minimum in the visible and UV, and a strong correlation of spectral slope between UV and visible wavelengths. These results do not support the reversal of the spectral slope in the UV relative to the visible wavelengths on a global scale (Hendrix et al., 2003). If space weathering causes spectral slope reversal from UV to visible (Hendrix and Vilas, 2006), then the lack of such a phenomenon on Vesta suggests the lack of global-scale space weathering, as previously inferred from visible and NIR data (e.g., Pieters et al., 2005).

In the range of 200 to 1000 nm, our rotationally averaged spectrum of Vesta can be well explained by a single-component, fine-grain howardite, or a linear mixing model with a small fraction (<20%) of fine-grain diogenite added. This is consistent with previous conclusions about the global surface composition of Vesta (Gaffey, 1997).

Finally, we want to point out that the main conclusions of this research rely on the combination of UV and visible spectral regions. The diversity of the spectral properties of asteroids in the UV suggests that this spectral range could reveal substantial clues to the surface compositions. It is necessary to systematically collect observational data of asteroids in the UV and to expand the spectral library of laboratory UV measurements.

**Acknowledgements**

Support for this work was provided by the National Aeronautics and Space Administration (NASA) through Grant HST-GO-12049.01-A from the Space Telescope Science Institute, which is operated by the Association of Universities for Research in Astronomy, Inc., under NASA Contract NAS5-26555. We thank the Swift team for the careful and successful planning of our observations and acknowledge support from the Swift Guest Investigator program. A portion of this work was carried out at the Jet Propulsion Laboratory, California

Institute of Technology, under contract with NASA. Select reflectance spectra were acquired from the RELAB database maintained by Brown University. RELAB is a multiuser facility operated under NASA Grant NAGW-748. JYL would like to extend special thanks to Dr. Amanda Hendrix and Dr. Faith Vilas for the very helpful discussions on their previous work. The authors are grateful to the two reviewers for their critical readings of this manuscript that have helped us clarify our conclusions and improve the manuscript substantially.

**Table 1.** List of HST/WFC3 and Swift/UVOT observations

| Date | Filter | Exposure Time (s) | Heliocentric Distance (AU) | Earth Range (AU) | Phase Angle (º) | Sub-Earth Latitude (º) | Sub-Solar Latitude (º) | Energy enclosure at 0.8" (radius) | $F_\odot$ (W/m²/nm) | Equivalent wavelength (nm) | 1-σ Bandwidth (nm) | Average geometric albedo |
|---|---|---|---|---|---|---|---|---|---|---|---|---|
| HST WFC3 Observations | | | | | | | | | | | | |
| Feb 25, 2010 | F225W | 45 | 2.39 | 1.42 | 5.3 | 23 | 27 | 92.5% | 0.07240 | 252.5 | -22.7, +19.8 | 0.13 |
| Feb 25, 2010 | F275W | 10 | | | | | | 92.3% | 0.2357 | 282.3 | -19.5, +14.1 | 0.16 |
| Feb 25, 2010 | F336W | 1.7 | | | | | | 92.8% | 0.8720 | 337.7 | -16.9, +17.1 | 0.22 |
| Feb 25/28, 2010 | F373N | 12 | | | 5.3, 6.4 | | | 93.1% | 1.032 | 373.0 | ±0.9 | 0.24 |
| Feb 25/28, 2010 | F469N | 4 | | | | | | 93.7% | 1.970 | 468.8 | ±1.0 | 0.34 |
| Feb 25/28, 2010 | F673N | 1.2 | | | | | | 94.4% | 1.505 | 676.6 | ±2.1 | 0.39 |
| Feb 25/28, 2010 | F953N | 12 | | | | | | 94.1% | 0.8226 | 953.5 | ±3.6 | 0.31 |
| Swift Observations | | | | | | | | | | | | |
| Apr 5/20, 2011 | UVM2 | ~128 | 2.16, 2.17 | 2.22, 2.05 | 26.4, 27.4 | -30, -31 | -20, -22 | N/A | 0.04622 | 244.9 | -20.9, +30.1 | 0.094 |
| | UVW2 | 13 – 68 | | | | | | | 0.08364 | 283.9 | -61.9, +35.1 | 0.11 |
| | Grism | ~700 | | | | | | | N/A | N/A | N/A | |

**Figure captions**

Fig 1 - Swift grism spectra of Vesta at three sub-Earth longitudes, compared with a rotationally averaged IUE spectrum. The error bars are point-to-point uncertainties due to photon noise. The spike at ~215 nm in the spectrum of 68º longitude is due to a background star and should be ignored. The down-turn in all Swift spectra longward of ~290 nm is due to coincident loss caused by the brightness of Vesta. Overall, Swift spectra are similar in shape to the IUE spectrum in the overlapping wavelengths. Swift systematically measured lower geometric albedos than IUE. This is most likely related to the systematic uncertainty of the two instruments, and is within the photometric calibration uncertainty of 25% for Swift.

Fig 2 - Panel (a) shows the UV-visible geometric albedo spectrum of Vesta from HST/WFC3, Swift/UVOT wideband photometry and grism spectroscopy, IUE, and SMASS II. Panel (b) amplifies the UV region. The IUE data are averaged over a full rotational lightcurve, and the Swift data plotted here corresponds to lightcurve midpoint. The SMASS II spectrum is scaled to match HST measurement at 673 nm. All other geometric albedo measurements are photometrically calibrated and not scaled. The y-error bars of photometric data points represent the photometric calibration uncertainties. The y-error bars of HST points in 253 nm, 282 nm, and 338 nm also include the estimated lightcurve amplitude of ~10%, with our data points near the minima. The x-error bars of photometric points represent the 1-$\sigma$ bandwidths as described in the text. The y-error bars of spectra represent the relative uncertainties caused by photon noise. The absolute photometric uncertainties of the Swift spectrum is ~25%, and the IUE spectrum ~15-25%. Structures in the UV spectra from both Swift and IUE are probably spectral features from the reference solar spectrum rather than real absorption features in Vesta's spectrum.

Fig 3 - Panel (a) shows the rotational lightcurves of Vesta at various wavelengths from HST/WFC3 and Swift/UVOT broadband photometry, plotted with respect to sub-Earth (east-)longitude. The filters used are marked on the right-hand side of the plot. The three filled triangles mark the longitudes of the Swift grism spectra shown in Fig. 3. Panel (b) shows the rotational lightcurves of Vesta in three wavelength regions extracted from IUE spectra. The lightcurve amplitude of Vesta remains ~10% from UV to visible wavelengths.

Fig 4 - UV spectral slope of Vesta between 240 nm and 320 nm measured from IUE spectra, plotted with respect to sub-Earth longitude. The slope is expressed as percentage per 100 nm at 280 nm. The error bars represent the fitted 1-$\sigma$ uncertainties of the slopes. The dip near 20º, the relatively high slope in the eastern hemisphere and moderately high slope in the western hemisphere all appear to be real and correspond to the spectrally blue, red, and moderately red spectral slopes in the visible wavelengths at similar longitudes.

Fig 5 - Panel (a) shows the comparison between the Vesta spectrum from 220 nm to 953 nm and the spectra of howardite and diogenite with various grain size. The small grain howardite fits the spectrum of Vesta the best. Panel (b) shows the various mixing models to fit the UV-visible spectrum of Vesta. The spectrum of linear mixing model with 80% small grain howardite and 20% small grain diogenite is similar to a single component small grain howardite, and fits Vesta's spectrum equally well. While the models with large grain components (Hendrix et al., 2003) fit the UV part well, they do not fit the 1-μm absorption or the spectral slope from 450 nm to 650 nm, likely due to the large grain size of laboratory measured spectra.

Fig. 1

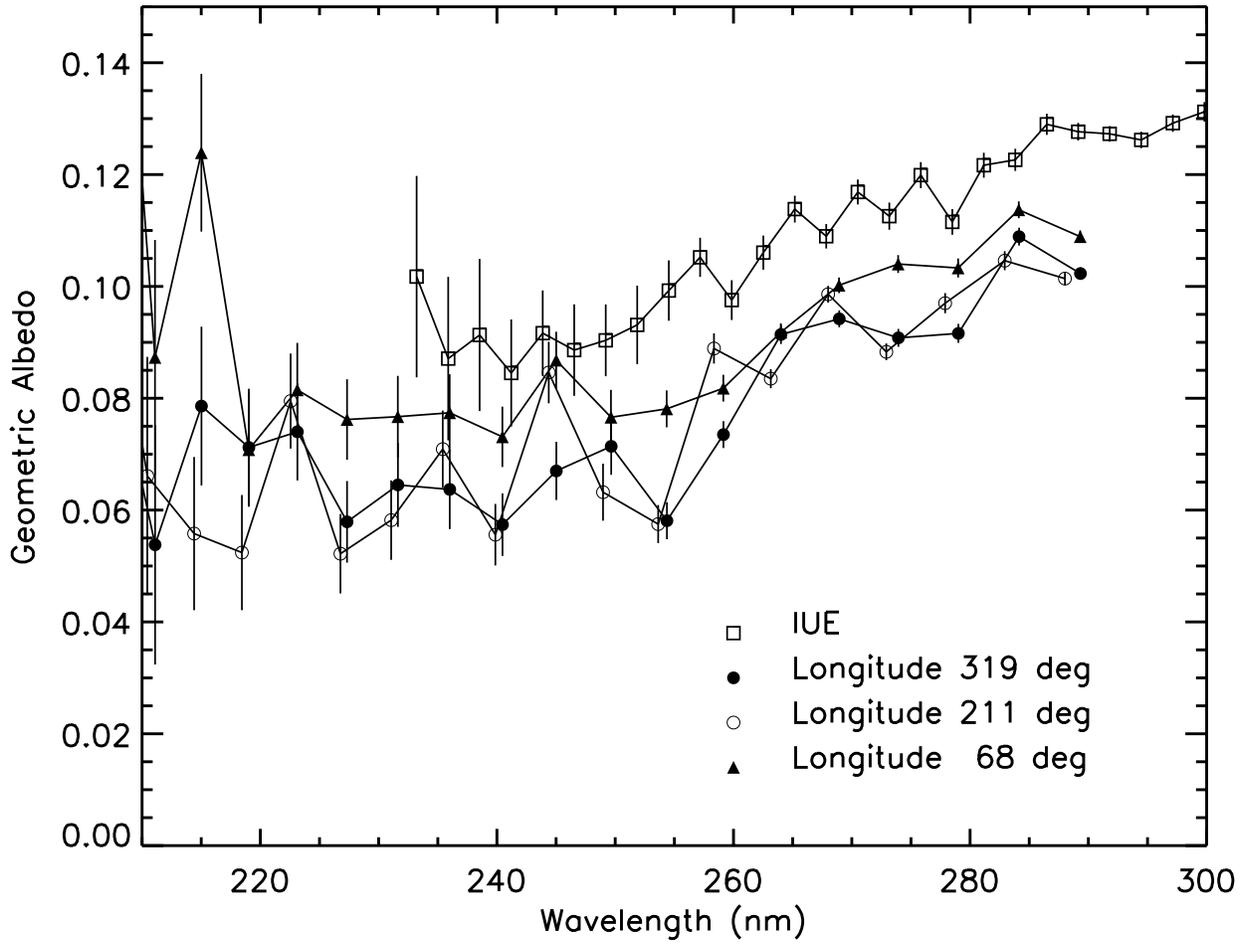

1 **Fig. 2**

2 **(a)**

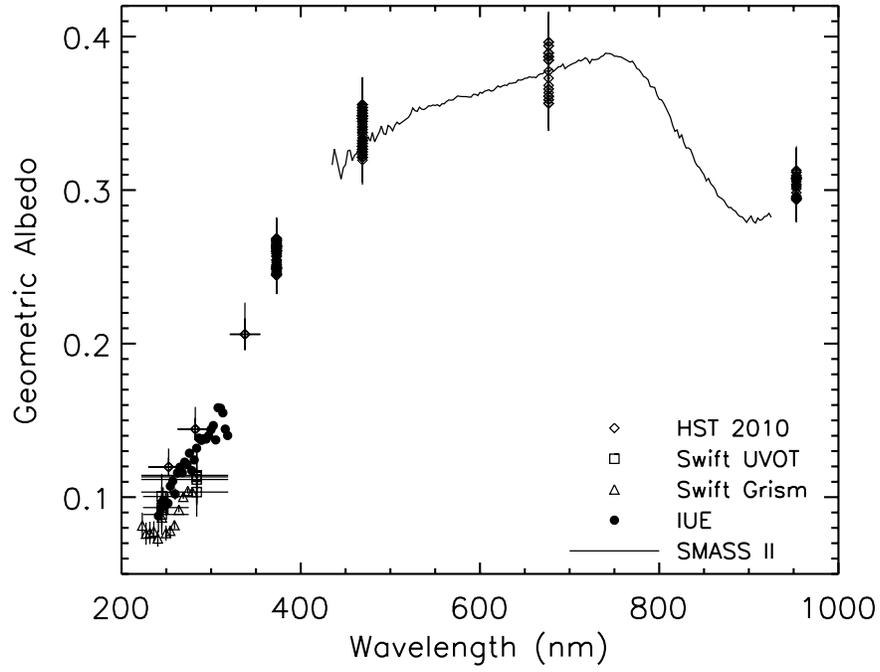

3 **(b)**

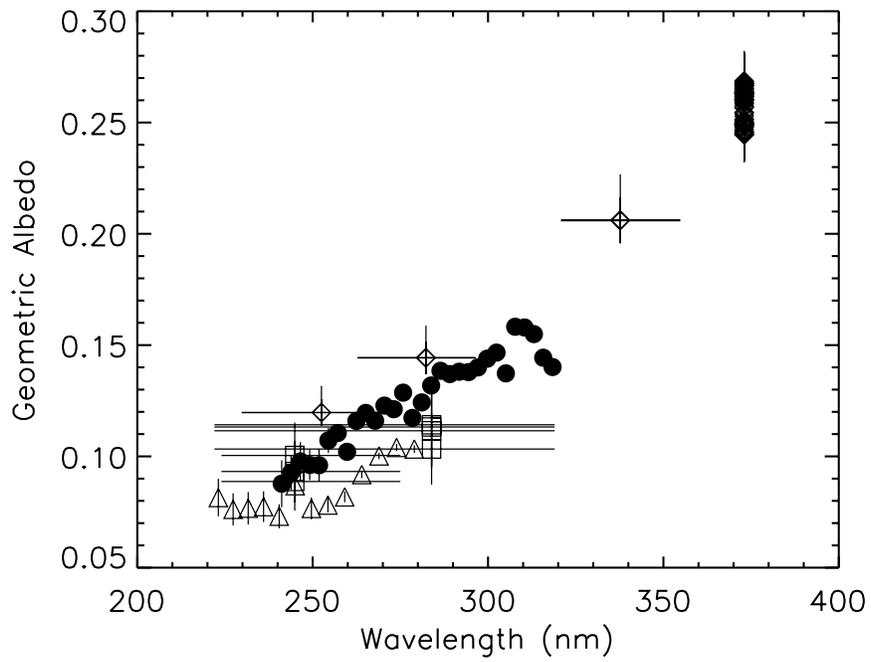

Fig. 3a

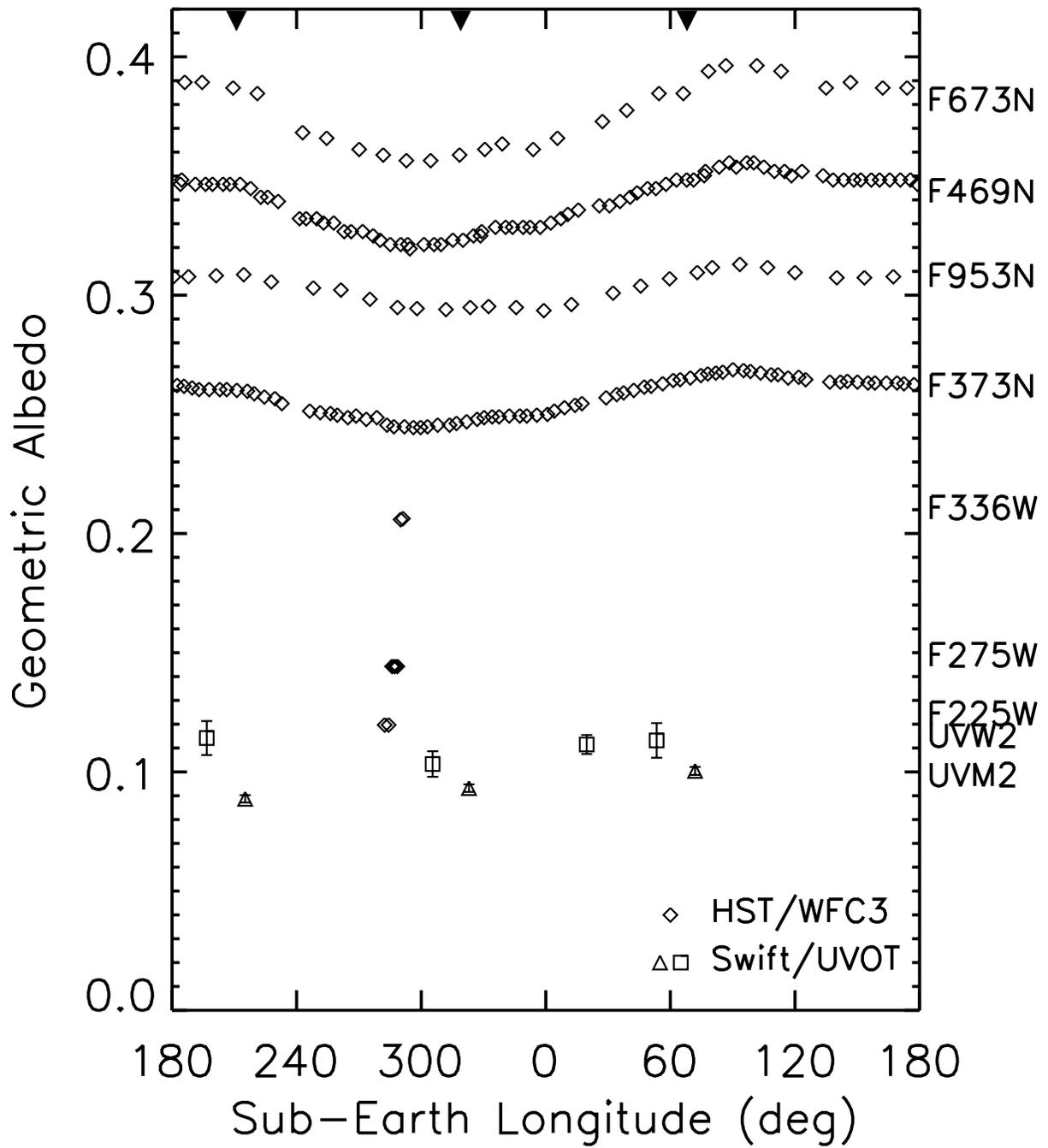

Fig. 3b

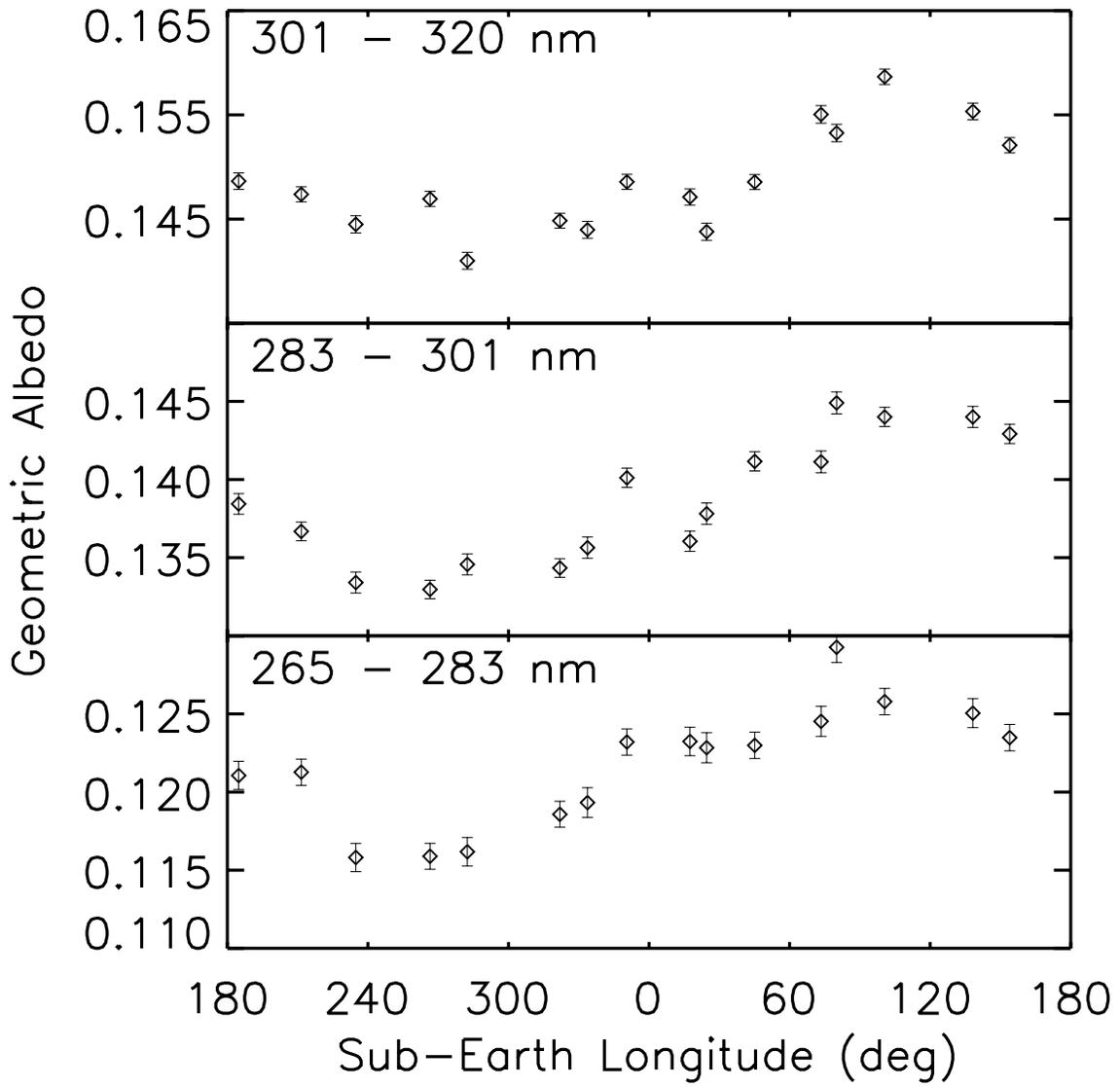

1 **Fig. 4**
2
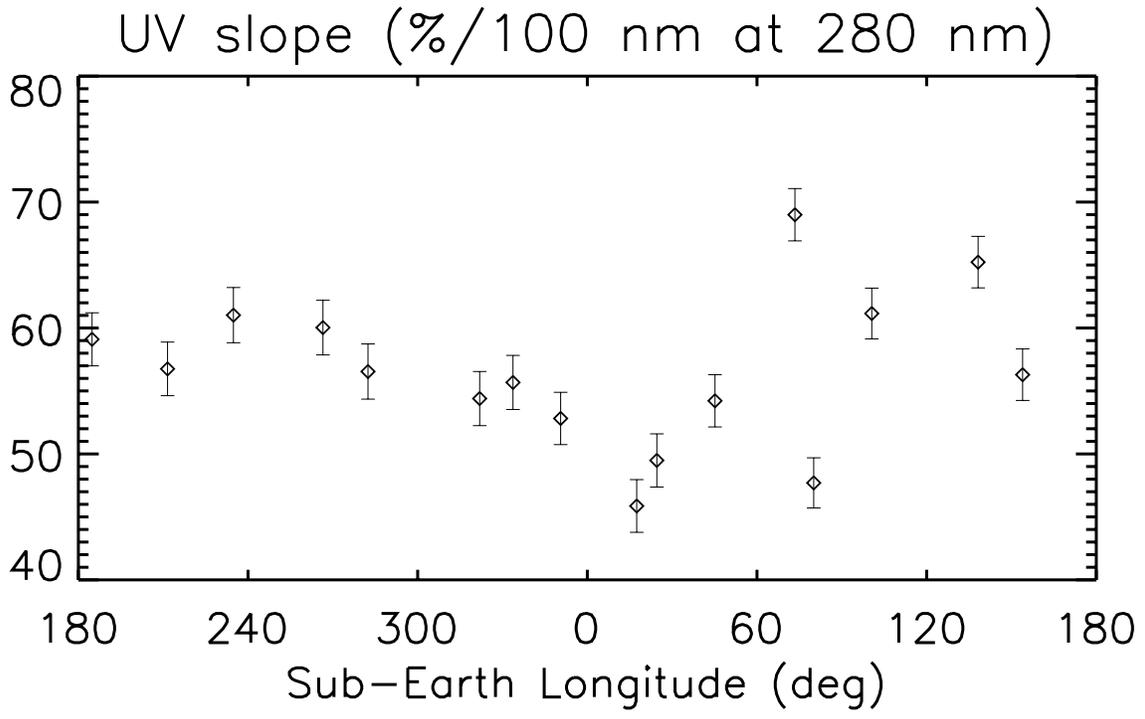
3

Fig. 5

(a)

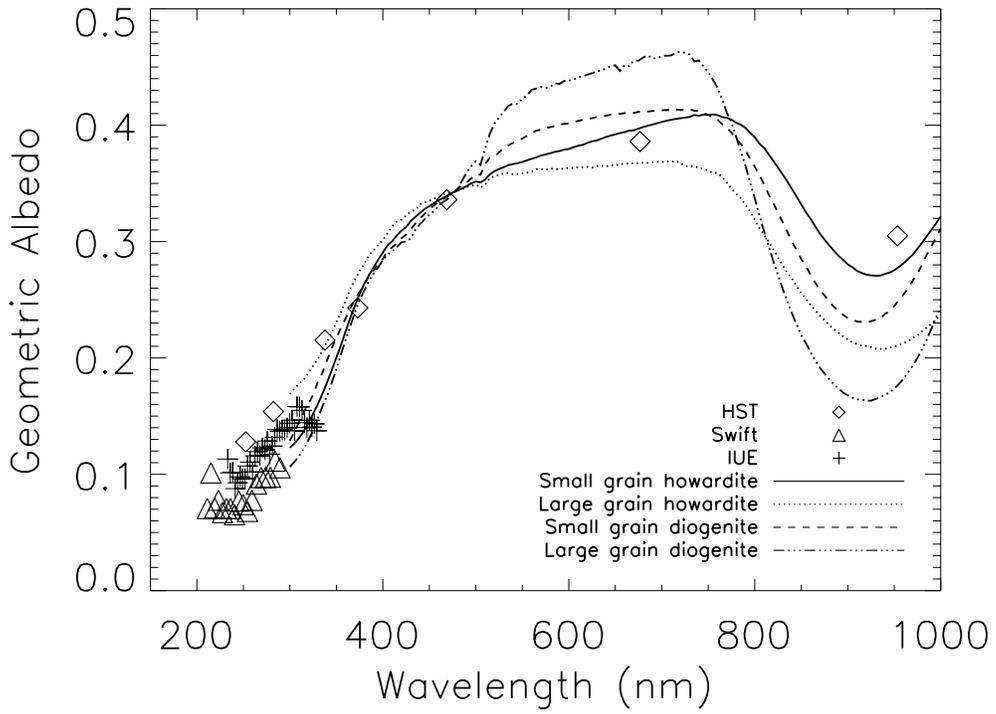

(b)

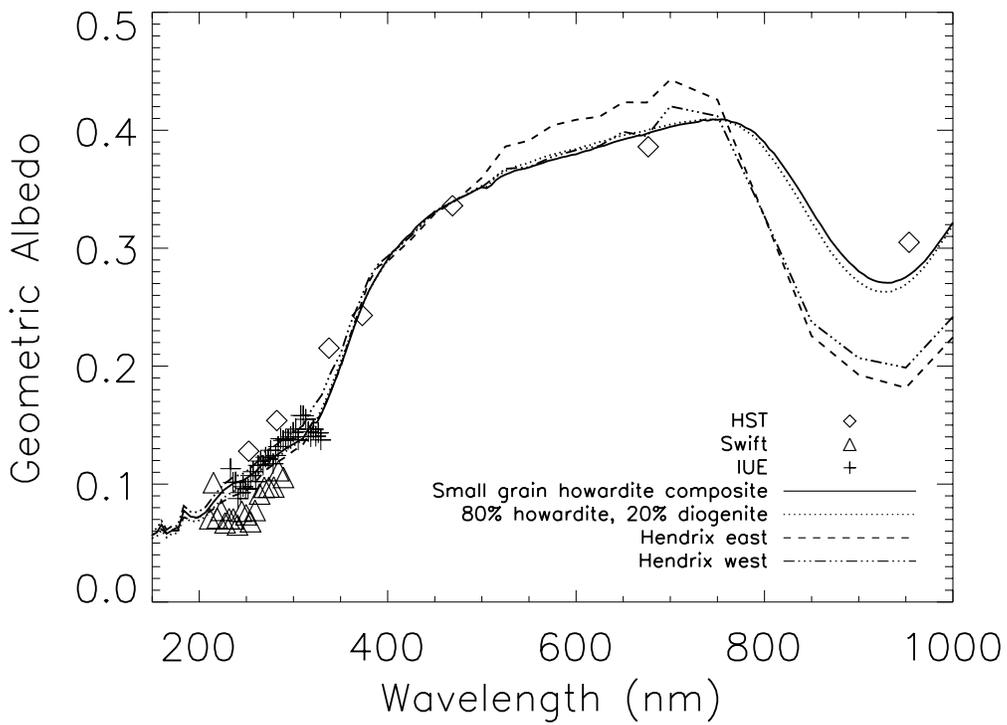